\documentstyle[epsfig,prl,twocolumn,eqsecnum,aps,floats]{revtex}
\begin{document}
\draft

\twocolumn[\hsize\textwidth\columnwidth\hsize\csname @twocolumnfalse\endcsname

\title{Free massive particles with total energy $E < mc^2$
       in curved spacetimes}

\author{Jorge Casti\~neiras\\
        Instituto de F\'\i sica Te\'orica, Universidade Estadual Paulista, 
        Rua Pamplona 145, 01405-900, S\~ao Paulo, S\~ao Paulo, Brazil}
\author{Luis C. B. Crispino\\
        Departamento de F\'\i sica, Universidade Federal do Par\'a,
        Campus Universit\'ario do Guam\'a, 66075-900, Bel\'em, Par\'a, Brazil}
\author{George E. A. Matsas\\
        Instituto de F\'\i sica Te\'orica, Universidade Estadual Paulista, 
        Rua Pamplona 145, 01405-900, S\~ao Paulo, S\~ao Paulo, Brazil}
\author{Daniel A. T. Vanzella\\
        Physics Department, University of Wisconsin-Milwaukee, 
        1900 E Kenwood Blvd., Milwaukee, WI 53211}

\maketitle

\hfuzz=25pt
\begin{abstract} 

We analyze {\em free elementary particles} with rest mass $m$ and {\em total}
energy $E < m c^2$ in  the Rindler wedge, outside Reissner-Nordstrom
black holes and  in the spacetime of relativistic (and non-relativistic) 
stars, and use Unruh-DeWitt-like detectors to calculate the associated 
particle detection rate  in each case. The (mean) particle position is 
identified with the spatial average of the excitation probability of the 
detectors, which are supposed to cover the whole space. Our results are 
shown to be in harmony with General Relativity classical predictions. 
Eventually we reconcile our conclusions with Earth-based experiments which 
are in good agreement with $E \geq m c^2$. 

\end{abstract}


\vskip2pc]

\section{Introduction}
\label{intro}

\narrowtext

The standard theory of quantum fields uses the fact that the Minkowski
spacetime is maximally symmetric.  The linear three momenta 
$(k^x,k^y,k^z)$ associated with the translational isometries on the 
spacelike hypersurfaces $t= {\rm const}$ constitutes a suitable set of quantum
numbers to label free particles, where we are assuming here that $(t,x,y,z)$
are the usual Minkowski coordinates.  In this simple case, the dispersion
relation $E \equiv \hbar \omega = \sqrt{| {\bf k} \, c |^2 + m^2 \, c^4 \;}$
imposes a simple constraint between the particle mass $m$, momentum ${\bf k}$
and energy $E$, and, thus, free particles with well defined linear momenta must
have total energy $E \ge m\, c^2$.  Because Earth-based particle experiments
assume in general the detection of asymptotic free states in Minkowski
spacetime, the possibility of measuring particles with total energy

\begin{equation} 
E < m c^2 
\label{fundamental} 
\end{equation} 
is usually neglected. Moreover, in the classical context of General 
Relativity, the detection {\em in loco} of {\em point} particles satisfying 
Eq.~(\ref{fundamental}) by direct capture is ruled out by the fact that 
an observer with four-velocity $u^\mu$ intercepting a particle with
four-momentum $k^\mu = m v^\mu$ ascribes to the particle an energy
$E = m v^\mu u_\mu \ge mc^2$. On the other hand, it is well known that,
unlike standard Quantum Field Theory in flat spacetime, the field quantization
carried over curved (stationary) backgrounds does not lead in general to 
any dispersion  relation for the energy and other quantum numbers, avoiding
thus the flat spacetime constraint $E \ge mc^2$. This can be understood
by recalling that the concept of {\em point particle} finds no room
in the context of Quantum Field Theory. Two main questions are raised, then, 
by the discussion above: Given a stationary spacetime 
{\em{\rm (i)} what is the minimum energy 
$E_{\rm min} \equiv \hbar \omega_{\rm min}$ 
allowed for a particle?} 
and 
{\em {\rm (ii)} what is the probability density associated with the detection
of  particles with $E\in [E_{\rm min}, m c^2)$ at different space points?}  

We analyze the questions above in the context of Quantum Theory of Linear
Fields in Curved
Spacetimes where the normal modes associated with our particles are seen as
free.  In order to define clearly what we mean by ``observing a particle'' we
use Unruh-DeWitt detectors (endowed with an internal structure defined by a
density of states). Our results are shown to be in agreement with General
Relativity predictions concerning the position of particles with $E< m c^2$
as defined by the same fiducial observers who proceed to the field 
quantization and with Earth-based experiments which usually assume 
$E \geq mc^2$.

The paper is organized as follows.  In Section~\ref{Rw} we consider a
two-dimensional Schwarzschild black hole which maintains a close 
relationship with
the Rindler wedge where a full analytical investigation can be carried out.
This section is important as a ``theoretical laboratory'' for the next ones and
to provide a better understanding of some phenomena as, e.g., the decay of
uniformly accelerated protons~\cite{MV1}.  This is so because the proton
interaction with {\em massive} particles possessing energy $E< m c^2$ plays an
important r\^ole according to coaccelerated observers with the proton.  In
Section~\ref{RN} we investigate the questions (i) and (ii) outside
Reissner-Nordstrom black holes.  We discuss, in particular, the agreement of
our results with General Relativity.  In Section~\ref{RS} we consider
relativistic (and non-relativistic) stars.  This case is qualitatively
different from the earlier ones because the energy spectrum of particles with
$E<m c^2$ is discrete.  This fact is a reflection of the nonexistence of event
horizons in the star spacetime.  We show eventually that for Earth-based
experiments $E_{\rm min} \approx m c^2$, as expected.  Our final remarks are
made in Section~\ref{conclusions}.  We will assume hereafter natural units,
$\hbar=c=G=1$, unless stated otherwise.

\section{Detection of massive particles in a 
two-dimensional black hole-like spacetime}
\label{Rw}

Let us begin considering the line element of a
two-dimensional Schwarzschild spacetime:

\begin{equation}
ds^2= \left( 1 - 2M/r \right) dt^2 - \left( 1 - 2M/r \right)^{-1} dr^2 \;.
\label{SS}
\end{equation}
Eq.~(\ref{SS}) can be seen as describing a two-dimensional black hole with mass
$M$.  Close to the horizon, $r \approx 2M$, Eq.~(\ref{SS}) can be written as

\begin{equation}
ds^2=  (\rho/4M)^2 dt^2 -  d\rho^2 \;,
\label{RW}
\end{equation}
where $\rho(r) \equiv \sqrt{8M(r-2M)}$.  (Note that in these coordinates the
horizon is at $\rho=0$.)  Line element~(\ref{RW}) is associated with the 
Rindler wedge (which is a globally hyperbolic spacetime) provided that 
$0<\rho<+\infty$ and $-\infty < t < +\infty$. The advantage of considering
spacetime~(\ref{RW}) rather than~(\ref{SS}) is three-fold: first it has the 
main relevant features (for our purposes) of the two-dimensional Schwarzschild
spacetime  (the fact that they differ asymptotically will not be important at
this point), second it provides a better understanding of some phenomena
occurring in uniformly accelerated frames and third it
allows a completely analytic discussion. 
\begin{figure}
\begin{center}
\mbox{\epsfig{file=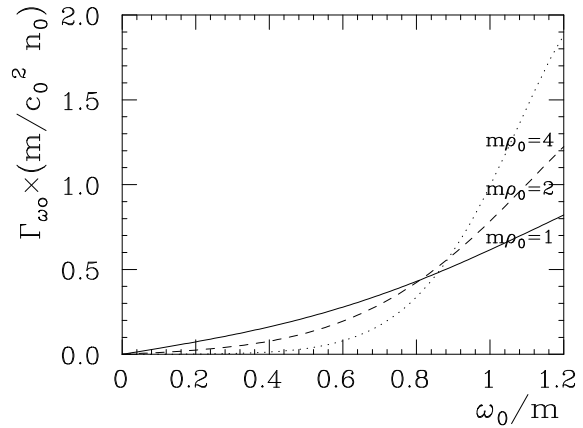,width=0.36 \textwidth,angle=90}}
\end{center}
\caption{We plot the detection rate ${\Gamma}_{\omega_0}$ as a function
of the  $\omega_0/m$ ratio for  observers  at different points 
$\rho_0$. For  observers far away from the horizon, there is a strong 
detection damping. 
}   
\label{fig1ccmv}
\end{figure}

Let us now consider in this spacetime a free scalar field $\hat \Phi(x)$ with
mass $m$. The positive-frequency solutions of the Klein-Gordon equation
$ ( \Box + m^2) u_\omega(x^\mu) = 0 $ can be written as 
$u_\omega(x^\mu) \equiv \psi_\omega (\rho) e^{-i \omega t}$
where
\begin{equation}
- \frac{d^2 \psi_\omega }{dx^2}
+V^{\rm RW}_{\rm eff} \psi_\omega (\rho) = \omega^2 \psi_\omega (\rho)\;.
\label{VRW}
\end{equation}
Here $x=4M \ln (\rho/4M) $ and
$V^{\rm RW}_{\rm eff} = (4M)^{-2} \rho^2 m^2 $. 
Note that the effective potential grows unboundedly at the infinity. 
By solving Eq.~(\ref{VRW}), we obtain
\begin{equation}
u_\omega (x^\mu) = \sqrt{(4M/\pi^2)\sinh (4 \pi M \omega)\;} 
               K_{4 i M \omega} (m \rho) e^{-i\omega t} \; ,
\label{u}
\end{equation}
where $K_\nu (x) $ is the modified Bessel function
and we have orthonormalized $u_\omega (x)$ according to the 
Klein-Gordon inner product~\cite{Fr}-\cite{F}. Note that $\omega \in (0,
+\infty)$, i.e.,  there are massive Rindler particles with arbitrarily small
energies.  

Next, the field is expanded in terms of positive- and negative-frequency
modes as usually:
\begin{equation}
\hat \Phi(x) 
=
\int_0^{+\infty} d\omega \;[ \hat a_\omega u_\omega (x) + H.c.] \;,
\label{Phi}
\end{equation}
where the annihilation and creation operators satisfy  
$[\hat a_\omega, \hat a^\dagger_{\omega'}] =
\delta(\omega- \omega').$ Since ${\partial_t} u_\omega = - i \omega
u_\omega$,  the {\em fiducial observer} with respect to whom the
quantization is performed is the one at $\rho = \rho_0 = 4 M$, whose proper 
time is $t$ [see Eq.~(\ref{RW})].  The Rindler vacuum $|0 \rangle$ is defined 
from $\hat a_\omega |0 \rangle =0$.

Now, we introduce an Unruh-DeWitt detector~\cite{UDW} described by a
localized monopole $\hat m(s)$ with proper time $s$ and worldline 
$z^\mu = z^\mu (s)$.  Let $\hat H$ be the detector free Hamiltonian acting 
as $\hat H |E \rangle = E |E \rangle$ on the detector energy eigenstates 
$|E \rangle$ and $\hat m(s) = e^{i \hat H s } \hat m(0) e^{-i \hat H s }$.  
We will denote by
$|E_G \rangle$ the detector ground state and assume $E_G \equiv 0$.  
The excited
states of the detector will be ruled by some (normalized) density of states
$\beta_{{E_0}} (E)$ peaked at $ E = {E_0}$ and satisfying
\begin{equation}
\int_{0}^{+\infty} d E \, \beta_{E_0} (E) = 1 \,.
\label{delta}
\end{equation}
The simplest choice is $ \beta_{E_0} (E) = \delta (E- {E_0})$
which characterizes a single excited-state detector. This will be enough 
in the black hole cases but not in the relativistic star one
where we have to deal with particles with discrete energy spectra. 
For the later case it is more convenient to consider 
\begin{eqnarray}
\beta_{E_0} (E) &=& ({n}/{E_0}) 
                  \left[ \Theta \left(E-E_0 + {E_0}/{2n} \right) \right.
\nonumber \\
                &-& \left. \Theta \left(E-E_0 - {E_0}/{2n} \right) 
                  \right] \; ,
\label{beta}
\end{eqnarray}
where $n = {\rm const.} \gg 1$ and $\Theta (x)$ is the step function.
We note that Eq.~(\ref{beta}) satisfies Eq.~(\ref{delta}) and the 
useful property $\beta_{c E_0} (c E) = c^{-1} \beta_{E_0} (E) $
for $c \in {\bf R}$.
Moreover, we can recover the single excited-state detector in the
$n \to +\infty$ limit, since 
$\lim_{n\to +\infty} \beta_{E_0} (E) = \delta (E- {E_0})$.
Once the detector is defined, we couple it to a massive scalar field
$\hat \Phi (x^\mu)$ through the interaction action
\begin{equation}
\hat S_I = 
\int_{-\infty}^{+\infty} ds \;c_0 \; \hat m(s)\; \hat \Phi [x^\mu(s)] \,,
\label{S}
\end{equation}
where $c_0$ is a small coupling constant. 

Let us now ask to our fiducial observer what is the total probability 
per (detector) proper time 
$\Gamma_\omega (\rho_d) \equiv {P}_\omega(\rho_d) / s^{\rm tot}_d$ of 
detecting a particle at some point $\rho_d$ with energy $\omega$. 
The excitation amplitude (at the tree level) 
$ {\cal A}_{\omega}^{\rm det} \equiv \langle 0
| \otimes \langle E | \hat S_I | E_G  \rangle \otimes | 
\omega \rangle $ associated with the particle detection  can be shown
to be
\begin{eqnarray}
{\cal A}_{\omega}^{\rm det} 
& = &  
4 \; c_0 \;  \sqrt{M \sinh(4 \pi M \omega)}\,  K_{4i M\omega}(m\rho_d)
\nonumber \\
&  \times & 
\, \delta(E- 4M\omega/\rho_d) \;,
\end{eqnarray}
where
the detector selectivity was chosen such that
$\langle E | m(0) | E_G  \rangle \equiv 1$.
The detection rate is thus given by
\begin{eqnarray}
&\Gamma_\omega& (\rho_d) 
= 
\frac{1}{s^{\rm tot}_d} \int_0^{+\infty} dE \;\beta_{E_0} (E) 
\int_0^{+\infty} d\omega' 
|{\cal A}_{\omega'}^{\rm det}|^2 F_\omega (\omega')
\nonumber \\
&=&
\frac{2 c_0^2}{\pi}  \sinh(\pi {E_0} \rho_d ) \rho_d
K_{i {E_0} \rho_d}^2 (m \rho_d ) 
F_\omega \left(\frac{{E_0} \rho_d}{4 M}\right) \; ,
\label{P2}
\end{eqnarray}
where we have chosen the density of states $\beta_{E_0} (E) = \delta (E-{E_0})$
and $F_\omega (\omega') $ characterizes a mixed state peaked at the particle
energy $\omega$ with the property that $F_\omega(\omega) = n_0 = {\rm const.}$ 
for every $\omega$. (One can avoid the introduction of mixed states by 
considering
the detection of wave packets, as shown below.)  Now, we carefully adjust the
detector at each point $\rho_d$ to {\em maximize} the detection 
probability~(\ref{P2}).
This is achieved by properly tuning its energy excitation gap:
$
{E_0} = 4M\omega/ \rho_d 
$
(note that ${E_0}$ and $\omega$ are related by a red-shift factor, 
as it should be). Hence Eq.~(\ref{P2}) becomes
\begin{equation}
 {\Gamma}_\omega (\rho_d) 
= 
\frac{2 c_0^2}{\pi}  n_0 \sinh(4\pi M \omega ) \rho_d
K_{4 i M \omega}^2 (m \rho_d ) \; .
\label{P2'}
\end{equation} 

Now, we may wonder what is
the probability rate ${\Gamma}_{\omega_0}$ that a massive state 
$|\omega_0 \rangle$ be
measured by our observer at $\rho_0$ in the particular case where the detector
is carried by our experimentalist, i.e.,  $\rho_d = \rho_0$
(and we recall that in the Rindler wedge $M = \rho_0/4$).  
In Fig.~(\ref{fig1ccmv}) we plot 
\begin{equation}
{\Gamma}_{\omega_0} 
=  
\frac{2 c_0^2}{\pi}  n_0 \sinh (\pi \rho_0 \omega_0 ) \rho_0
K_{i \rho_0 \omega_0}^2 (m \rho_0 ) \; ,
\label{P2a'}
\end{equation}
as a function of $\omega_0/m$.  We clearly note 
that the farther away the observer is from the horizon,
the steeper the detection damping for $\omega_0/m < 1$.

\begin{figure}
\begin{center}
\mbox{\epsfig{file=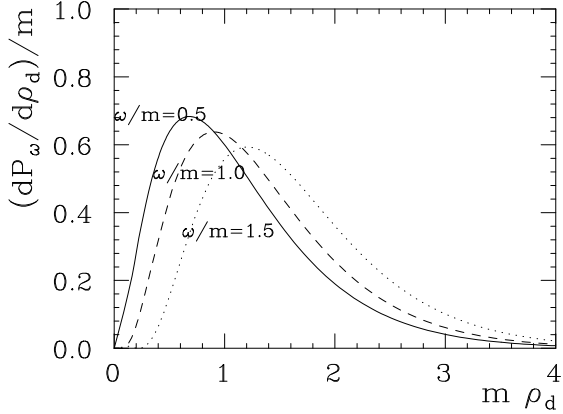,width=0.36\textwidth,angle=90}}
\end{center}
\caption{We plot the probability density $d{\cal P}_\omega/d\rho_d$ 
for different $\omega/m$ ratios, where we have assumed $Mm=1/4$. We 
note that the smaller the $\omega/m$ ratio,  the closer to the
horizon (in average) the particle lies, where the 
``gravitational potential'' is stronger.}
\label{fig2ccmv}
\end{figure}

Next, let us define from Eq.~(\ref{P2'}) the normalized probability density  
\begin{equation}
d{\cal P}_\omega/d\rho_d 
\equiv 
{\Gamma}_\omega (\rho_d)/\int_0^{+\infty}{\Gamma}_\omega (\rho'_d) d\rho'_d\; .
\label{density}
\end{equation}
$ (d{\cal P}_\omega/d\rho_d) d\rho_d$ 
is the probability that a particle with energy 
$\omega$ be found between $\rho_d$ and $\rho_d + d\rho_d$.
We see from Fig.~(\ref{fig2ccmv}) that observers far away from 
the horizon will only be able to interact  with the ``tail'' of 
the ``wave functions'' associated with particles with small $\omega/m$.

The physical content carried out by 
Eq.~(\ref{density}) can be reproduced by considering
wave packets (rather than mixed states) as follows.
Let us represent a particle with typical energy $\omega$
through a wave packet defined as 
\begin{equation}
|\phi_\omega \rangle \equiv
\int_0^{+\infty} d\omega' G_\omega(\omega' ) \; 
\hat a^\dagger_{\omega'} |0 \rangle \; ,
\label{packet}
\end{equation} 
where $G_\omega(\omega' )$ 
is a peaked function at $\omega' = \omega$ and
$
\int_0^{+\infty} d\omega' |G_\omega (\omega')|^2 = 1  
$
in order that $\langle \phi_\omega | \phi_\omega \rangle =1$.
The {\em total} probability for detecting $|\phi_\omega \rangle$ 
(at the tree level)
at some $\rho_d$ between the constant-coordinate-time hypersurfaces 
$t= {\rm const} \to -\infty$ and $t= {\rm const} \to +\infty$ is
$
\tilde {\cal P}^{wp}_\omega (\rho_d)
=
|\langle 0|\otimes \langle E_0 
|\hat S_I| 
E_G  \rangle \otimes| \phi_\omega \rangle|^2 
\; , 
$ 
where we should tune the detector as before, $ E_0 = 4M\omega/ \rho_d $,
to maximize its detection probability.
In order to obtain the probability between the constant-proper-time 
hypersurfaces  $s= {\rm const} \to -\infty$ and $s= {\rm const} \to+\infty$, 
we must multiply both sides by the red-shift factor $4M/\rho_d$:
\begin{equation}
{\cal P}^{wp}_\omega (\rho_d) = 4 c_0^2 |G_\omega (\omega)|^2
\sinh(4 \pi M \omega) 
\rho_d\; K^2_{4 i  M \omega }\left(m \rho_d  \right) \;,
\label{P2''}
\end{equation}
where  
$G_\omega(\omega) = {\rm const}$.  Note that Eqs.~(\ref{P2''}) and~(\ref{P2'}) 
differ only by a (dimensional) constant factor. Thus by replacing 
$\Gamma_\omega (\rho_d)$ by ${\cal P}^{wp}_\omega (\rho_d)$ in 
Eq.~(\ref{density}),   
we obtain the same probability density $d {\cal P}_\omega/d\rho_d$.

Now, in order to interpret Eq.~(\ref{density}) on General Relativity bases, 
let us first consider a row of detectors each of them
lying at different $\rho_d$ and define the {\em average detection
position}
\begin{equation}
\langle \rho_d \rangle 
\equiv 
\int_0^{+\infty} \; d\rho_d \;\rho_d \;d {\cal P}_\omega / d \rho_d \;.
\label{averagedef}
\end{equation} 
By using Eq.~(\ref{density}), we obtain
\begin{eqnarray}
\langle \rho_d \rangle & = & 
\frac{\pi\tanh (4\pi M \omega) (64 M^2 \omega^2 +1 ) }{64 m M \omega}  
\nonumber \\
& \approx &
{\pi \;M  \omega}/{ m } 
\;\;\;\;\;\;\;\;\;\;\;\;\;\;\;\;\;\;\; (\omega \gg a) \;,
\label{average}
\end{eqnarray} 
where $a \equiv 1/4M$ is the proper acceleration of the fiducial observer.

\begin{figure}
\begin{center}
\mbox{\epsfig{file=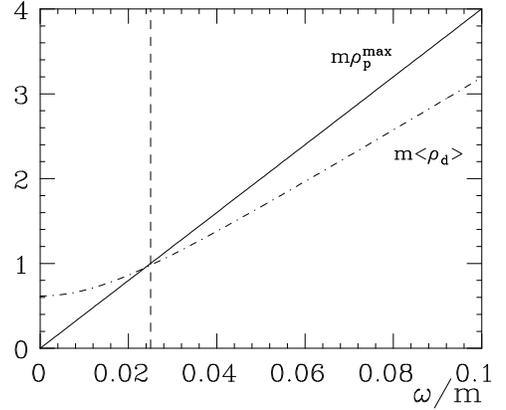,width=0.36\textwidth,angle=90}}
\end{center}
\caption{$\langle \rho_d \rangle$ is shown to be smaller than $\rho_p^{\rm max}
\equiv 4 M \omega / m$ {\em in the ``high-frequency'' regime 
$\omega > (4 M)^{-1} $} (i.e., at the right of the vertical broken line) 
just as expected. (We have assumed here $m M = 10$ but this nice agreement is
verified for any $m M$.)}
\label{fig3ccmv}
\end{figure}

Now, from General Relativity, a classical particle with mass $m$ 
{\em lying at rest} 
at some point $\rho_p$  has, according to our fiducial observer at  
$\rho_0 = 4M$, total energy $\omega = m \rho_p/4M$. By considering that 
the particle  may have in addition some kinetic energy, 
the total energy would be $\omega \ge m \rho_p/4M$. 
By inverting this equation, we obtain  
\begin{equation}
\rho_p   \le \rho_p^{\rm max} \equiv {4 M \omega}/{ m }   \;,
\label{particle}
\end{equation} 
which is expected to agree with $\langle \rho_d \rangle$, i.e., 
$\langle \rho_d \rangle \le  \rho_p^{\rm max}$, at least
in the ``high-frequency'' regime $\omega \gg a$ (where the 
quantum and classical behaviors may be compared). 
This conclusion is indeed in agreement
with Eqs.~(\ref{average}) and (\ref{particle})  
(see also Fig.~\ref{fig3ccmv}). The smaller the $\omega/m$ ratio, the
more likely to detect the particle closer to the horizon
where the ``effective gravitational potential'' decreases its total energy. 

\section{Detection of massive particles outside Reissner-Nordstrom black holes}
\label{RN}

Let us consider now the line element of a globally-hyperbolic
spherically-symmetric static spacetime  
\begin{equation}
ds^2 =  f(r) dt^2 - h(r) dr^2 - r^2 (d\theta^2 + \sin^2 \theta d\varphi^2)\; ,
\label{lineelementspheric}
\end{equation}
where we assume that we are restricted to a region ($f(r)>0$) 
where the Killing field $(\partial_t)^\mu$ is {\em timelike}.
We may cast the positive-frequency solutions of the Klein-Gordon 
equation 
$
( \Box + m^2)  u^\alpha_{\omega l {\rm m}}(x) =0
$
in the form ($l \in {\bf N}$, $-l \leq {\rm m} \leq l$)
\begin{equation}
u^\alpha_{\omega l {\rm m}}(x) = 
\sqrt{\frac{\omega}{\pi}} \frac{\psi^\alpha_{\omega l}(r)}{r} 
Y_{l {\rm m}} (\theta, \varphi) e^{-i \omega t} \;\; ,
\label{normalmodespheric}
\end{equation}
where 
$
\psi^\alpha_{\omega l}(r)
$ 
satisfies
\begin{equation}
-\sqrt{\frac{f}{h}} \frac{d}{dr} \left( \sqrt{\frac{f}{h}} \frac{d}{dr}\right) 
\psi^\alpha_{\omega l}(r)
+  V_{\rm eff} \, \psi^\alpha_{\omega l}(r)
= \omega^2 \psi^\alpha_{\omega l}(r)
\label{radialequation}
\end{equation}
with the following scattering potential:
\begin{equation}
V_{\rm eff} \equiv
\left[ 
\frac{\sqrt{{f}/{h}}}{r} \frac{d\sqrt{{f}/{h}}}{dr}  + 
\frac{l(l+1)f}{r^2}+m^2 f
\right]\; .
\label{Veff}
\end{equation}
The label $\alpha$ was introduced to distinguish between the two 
independent solutions of the radial equation~(\ref{radialequation}).
\begin{figure}
\begin{center}
\mbox{\epsfig{file=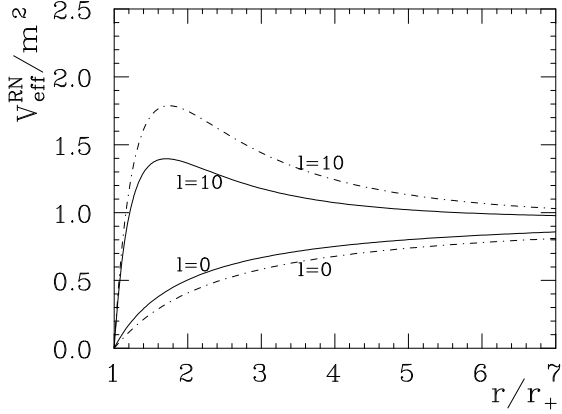,width=0.36\textwidth,angle=90}}
\end{center}
\caption{The scattering potential is plotted for $l=0$  
and $l=10$. Full and dashed lines are related with $Q=0$ and $Q=0.9 M$, 
respectively. $V^{\rm RN}_{\rm eff}$ vanishes
at the horizon and tends asymptotically to $m^2$. We note that 
the larger the angular momentum, 
the higher the potential and the more difficult to find a particle 
with $\omega/m < 1$ far from the horizon. (We have assumed here $mM=2$.)}   
\label{Vl}
\end{figure}

Outside a static black hole ($r>r_+$) with mass $M$ and 
charge $Q$ ($Q\leq M$), we have $f^{\rm RN}(r)=1/ h^{\rm RN}(r)$, 
where
\begin{equation}
f^{\rm RN}(r) = (1-r_+/r)(1-r_-/r) > 0
\label{g00}
\end{equation}
and 
$r_{\pm} \equiv M \pm\sqrt{M^2 - Q^2}$. (The label ``RN'' 
is a shortcut for Reissner-Nordstrom.)
In this case, the scattering potential
\begin{equation}
V_{\rm eff}^{\rm RN} = 
\left[1- \frac{2M}{r} + \frac{Q^2}{r^2}\right]
\left[ \frac{2M}{r^3} - \frac{2 Q^2}{r^4} + \frac{l(l+1)}{r^2} +
m^2 \right]  
\label{VeffRN}
\end{equation} 
vanishes at the horizon, $r=r_+$, and tends to $m^2$ asymptotically [see 
Fig.~(\ref{Vl})]. Thus, only ``{\em outgoing}'' particles, 
$\alpha \equiv \rightarrow$, from the white hole
horizon ${\cal H}^-$ will be able to comply 
Eq.~(\ref{fundamental}). (Eventually they will be totally reflected to the
black hole horizon ${\cal H}^+$.) {\em Incoming} particles from ${\cal J}^-$, 
$\alpha \equiv \leftarrow$, will necessarily have $\omega \ge m$.
[The case $\omega = m$ is somewhat subtler and will not be considered 
here (see Refs.~\cite{CM}-\cite{HMS2} for the case $\omega=m=0$).]
Let us note that 
the larger the $l$, the higher the centrifugal barrier of
the potential [see Fig.~(\ref{Vl})]. 
As a consequence, particles with
$l\gg 1$ which are found to verify Eq.~(\ref{fundamental}) will be necessarily
restricted to a close neighborhood  around the horizon, where the gravitational
field is strong enough to compensate the angular momentum effect.
Hence, in order to find particles satisfying Eq.~(\ref{fundamental}) 
``relatively'' far from the horizon, we will focus on particles with
$l=0$.

\begin{figure}
\begin{center}
\mbox{\epsfig{file=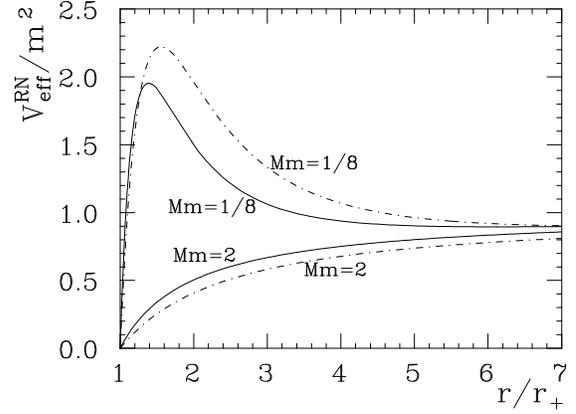,width=0.36\textwidth,angle=90}}
\end{center}
\caption{The scattering potential is plotted for $mM=1/8$ 
and $mM=2$. Full and dashed lines are related with $Q=0$ and $Q=0.9 M$,
respectively.
We note that the larger $mM$, the more favorable 
to detect particles with $\omega/m<1$. (Here we have assumed $l=0$.)
}
\label{VMm}
\end{figure}

For chargeless black holes, the local extrema of the 
scattering  potential can be analytically determined. 
For the case $l=0$ and $M m < 1/4$, we have 
\begin{equation}
r_1=
\frac{1}{m}\left( \sqrt{3\,} \sin \frac{\xi}{3} - \cos \frac{\xi}{3} \right)
\; {\rm and} \;\;
r_2=
\frac{2}{m} \cos \frac{\xi}{3}\,, 
\label{roots}
\end{equation}
where $r_1$ and $r_2$ are associated with a local 
maximum and minimum ($r_1 < r_2$), respectively, 
and $ \xi \equiv \pi -\arctan [\sqrt {1- 16 (mM)^2}/(4 mM)]$.
For the case $l=0$ and $M m > 1/4$ there are no 
local extrema [see Fig.~(\ref{VMm})]. 
For $M m = 1/4$, $r_1=r_2$ is an inflection point and  
$V_{\rm eff}^{\rm RN}|_{r=r_1,r_2} = 3 m^2 /4$. 
For a black hole with $M \geq 3 M_\odot$ 
and a particle with mass $m \geq m_{e^-}$,
we have $M m \geq 10^{16}$. This is why we will focus 
our simulations on values $M m > 1/4$. An analytic investigation of the
local extrema when $Q\neq 0$ is not possible since 
$d V_{\rm eff}^{\rm RN}/dr=0$ results in a full fifth-order 
algebraic equation but it can be seen numerically [see, e.g., 
Figs.~(\ref{Vl}) and~(\ref{VMm})] 
that  the presence of charge does not change significantly
the form of the potential 
$ V_{\rm eff}^{\rm RN} (r/r_+)$.

By using the coordinate transformation~\cite{CM}
\begin{equation}
r \to x = y + \frac{y_-^2 \ln (y-y_-) - y_+^2 \ln (y-y_+)}{y_- - y_+}\; ,
\label{rtox}
\end{equation}
where
$y\equiv r/2M$ and $y_\pm \equiv r_\pm/2M$, we can rewrite 
Eq.~(\ref{radialequation}) as 
\begin{equation}
-\frac{d^2 \psi^\alpha_{\omega l}}{d x^2} + 
4 M^2 V_{\rm eff}^{\rm RN} (x[y(r)]) \psi^\alpha_{\omega l} 
= 4 M^2 \omega^2 \psi^\alpha_{\omega l} \; .
\label{radialequation2}
\end{equation}
Thus, close to and far from the horizon, we
can write the outgoing $\psi^{\alpha}_{\omega l}$ functions  as 
($\omega \geq 0$, i.e., $\omega_{\rm min} = 0$)
\begin{equation}
\psi_{\omega l}^{\rightarrow}(x)
       \approx A^{\rightarrow}_{\omega l}
         \left\{
            \begin{array}{lc}
                e^{2iM\omega x} + {\cal R}_{\omega l}^{\rightarrow} 
                e^{-2iM\omega x}  & 
                (x < 0\, , |x| \gg 1) \\ 
                {\cal T}_{\omega l}^{\rightarrow} 
                e^{2iM \tilde \omega x} & (x \gg 1) 
          \end{array} 
        \right.   
   \label{51}
\end{equation}
and the incoming ones as ($\omega \ge m$, i.e., $\omega_{\rm min} = m $) 
\begin{equation}
    \psi_{\omega l}^{\leftarrow}(x) 
        \approx 
           A^{\leftarrow}_{\omega l}
             \left\{ 
               \begin{array}{lc} 
                  {\cal T}_{\omega l}^{\leftarrow} e^{-2iM\omega x} 
                  & (x< 0\, , |x| \gg 1) \\
                  e^{-2iM\tilde{\omega} x} + {\cal R}_{\omega l}^{\leftarrow}  
                  e^{2iM \tilde{\omega} x}
                  & (x \gg 1),
          \end{array} 
      \right. 
   \label{68}
\end{equation}
where $\tilde \omega \equiv \sqrt{\omega^2 - m^2}$.
For $\omega \ge m$,
$
\left| {\cal R}_{\omega l}^{\leftarrow} \right|^2 ,
\left| {\cal R}_{\omega l}^{\rightarrow} \right|^2 
$  
and
$
\left| {\cal T}_{\omega l}^{\leftarrow} \right|^2 ,
\left| {\cal T}_{\omega l}^{\rightarrow} \right|^2
$
can be seen as reflection and transmission coefficients, respectively, 
satisfying the probability conservation equations
\begin{equation}
\left| {\cal R}_{\omega l}^\rightarrow \right|^2 
+ 
\frac{\tilde \omega}{\omega}  
\left| {\cal T}_{\omega l}^\rightarrow \right|^2 
= 1 
\;\; 
{\rm and} 
\;\;
\left| {\cal R}_{\omega l}^\leftarrow \right|^2 
+ 
\frac{\omega}{\tilde \omega} 
\left| {\cal T}_{\omega l}^\leftarrow \right|^2
= 1. 
\label{conservation}
\end{equation}
For $\omega < m$, the outgoing modes fade exponentially far away from the
horizon and $ | {\cal R}_{\omega l}^\rightarrow| = 1 $.
One of the tests performed to guaranty the reliability of our codes 
was the verification that Eqs.~(\ref{51})-(\ref{68}) 
were satisfied 
(with the appropriate relations for 
${\cal R}_{\omega l}^\alpha$ 
and 
${\cal T}_{\omega l}^\alpha$) 
along the numerical calculation.
\begin{figure}
\begin{center}
\mbox{\epsfig{file=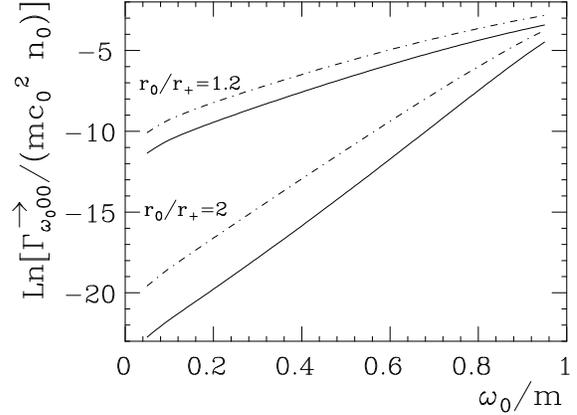,width=0.36 \textwidth,angle=90}}
\end{center}
\caption{We plot the detection rate ${\Gamma}^{\rightarrow}_{\omega_0 00}$ 
as a function
of the  $\omega_0/m$ ratio for  observers  at different points 
$r_0$. Full and dashed lines are related with $Q=0$ and $Q=0.9 M$, 
respectively. The damping in the detection rate is steeper for observers
far away from the horizon. (We have assumed here $mM=2$.)
}   
\label{fig5accmv}
\end{figure}

We normalize  $\psi_{\omega l}^{\alpha}$ such that 
$u_{\omega l {\rm m}}^{\alpha}$ are Klein-Gordon 
orthonormalized~\cite{Fr}-\cite{F}:
\begin{equation}
 i\int_{\Sigma_t} d\Sigma \; n^\mu 
 \left( 
 {u^\alpha_{\omega l {\rm m}}}^* \nabla_\mu u^{\alpha'}_{\omega' l' {\rm m}'}-
 u^{\alpha'}_{\omega' l' {\rm m}'} \nabla_\mu {u^\alpha_{\omega l {\rm m}}}^*  
 \right) 
     = 
     \delta_{A A'}
     \label{KG1} 
\end{equation}     
\begin{equation}   
   i\int_{\Sigma_t} d\Sigma \;  n^\mu
   \left( 
   u^\alpha_{\omega l {\rm m}} \nabla_\mu u^{\alpha'}_{\omega' l' {\rm m}'}
   - u^{\alpha'}_{\omega' l' {\rm m}'} \nabla_\mu u^\alpha_{\omega l {\rm m}}  
   \right) 
     = 0,
   \label{KG2}
\end{equation}
where here
$
\delta_{A A'} \equiv
     \delta_{\alpha \alpha'} \delta_{l l'} 
     \delta_{{\rm m} {\rm m}'} \delta(\omega - \omega')\; 
$
and $n^\mu$ is the future-pointing unit vector normal to the volume 
element of the  Cauchy surface $\Sigma_t$. By using 
Eq.~(\ref{radialequation2}) to transform the left-hand side of Eq.~(\ref{KG1})
in a surface term, we obtain the  normalization constants 
$A^\rightarrow_{\omega l} = (2\omega)^{-1} $ and  
$A^\leftarrow_{\omega l} = (2 \sqrt{\omega \tilde \omega})^{-1}$
(up to an arbitrary phase).
The incoming and outgoing modes are orthogonal to each other with respect to
the Klein-Gordon inner product~(\ref{KG1})-(\ref{KG2}). This can be  seen
by choosing $\Sigma_t ={\cal H}^- \cup {\cal J}^- $ 
in Eq.~(\ref{KG1})  and recalling that 
$\psi_{\omega l}^{\rightarrow}(x)$ 
and 
$\psi_{\omega l}^{\leftarrow}(x) $ 
vanish on $ {\cal J}^-$ and ${\cal H}^-$, respectively.
\begin{figure}
\begin{center}
\mbox{\epsfig{file=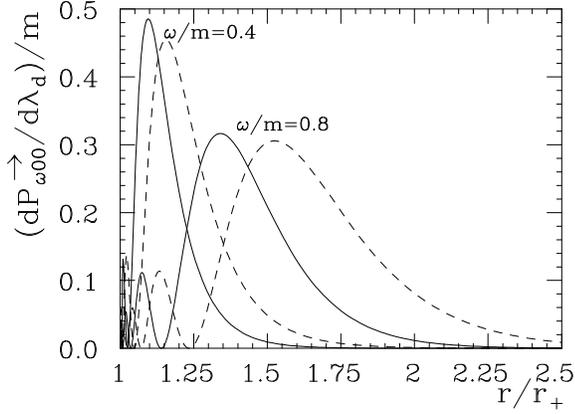,width=0.36\textwidth,angle=90}}
\end{center}
\caption{The probability density 
$(d{\cal P}^\rightarrow_{\omega 0 0}/d\lambda_d)/m$ is plotted 
for $\omega/m= 0.4$ and $\omega/m=0.8$.
Full and dashed lines are related with $Q=0$ and $Q=0.9 M$, respectively.
(Here we have assumed $Mm=2$.) 
We note that the smaller the $\omega/m$ ratio, the closer to the horizon
the particle should be.}
\label{excitationdensityfig}
\end{figure}

Next we expand a massive scalar field $\hat \Phi(x)$ in terms of positive
and negative frequency modes:
\begin{equation}
\hat \Phi(x) = 
\sum_{\alpha= \rightarrow}^\leftarrow
\sum_{l=0}^{+\infty} 
\sum_{{\rm m}=-l}^{l} 
\int_{\omega_{\rm min}}^{+\infty} d\omega
\;[ \hat a^\alpha_{\omega l {\rm m}} u^\alpha_{\omega l {\rm m}} (x) 
+ {\rm H.c.}] \;,
\label{Phisphericalsymmetric}
\end{equation}
where 
$\omega_{\rm min} = 0$ when $\alpha= \rightarrow$ 
and
$\omega_{\rm min} = m$ when $\alpha= \leftarrow$. 
As a consequence of orthonormalizing the normal modes with 
the  Klein-Gordon inner-product, 
the annihilation $\hat a^\alpha_{\omega l {\rm m}} $ and creation 
$\hat a^{\alpha \dagger}_{\omega l {\rm m}} $ operators  satisfy 
the simple commutation 
relations
\begin{equation}
     \left[ 
     \hat a^\alpha_{\omega l {\rm m}}, 
     {\hat a^{\alpha' \dagger}_{\omega' l' {\rm m}'}} 
     \right] = 
    \delta_{\alpha \alpha'} \delta_{l l'}
    \delta_{{\rm m} {\rm m}'}\delta(\omega - \omega')\;.
   \label{commutationrelation2}
\end{equation}
The Boulware vacuum $|0\rangle$ is defined by
$\hat a^\alpha_{\omega l {\rm m}}|0\rangle = 0$ for every 
$\alpha, \omega, l$ and ${\rm m}$.

Let us assume an Unruh-DeWitt detector coupled to the massive
scalar field as described by the interaction action~(\ref{S}).
The detection rate (detection probability per detector proper time) 
of particles with quantum numbers $\alpha, \omega, l$ and ${\rm m}$ 
(as defined by asymptotic fiducial observers) can be calculated (at the tree 
level):
\begin{equation}
{\Gamma}^\alpha_{\omega l {\rm m}} 
=    
2 c_0^2  n_0 \omega \sqrt{f^{\rm RN}( r_d )} 
\frac{|\psi^\alpha_{\omega l} ( r_d)|^2}{r_d^2} 
|Y_{l {\rm m}} (\theta_d, \varphi_d)|^2 \; ,
\label{P3}
\end{equation}
where $ n_0 = F_\omega (\omega) = {\rm const.} $, we have chosen 
$\beta_{E_0} (E) = \delta (E - E_0) $ and  we have tuned again 
the energy gap  of the detector at each point
$r_d$ in order to maximize the detection probability, 
namely, $E_0 = \omega/\sqrt{f^{\rm RN}(r_d)} $.

As in the case of the Rindler wedge, we may calculate the detection 
rate ${\Gamma}^{\alpha}_{\omega_0 l m}$ in the particular 
case where the massive state $|\alpha\, \omega_0\, l\, {\rm m}\rangle$ is
defined  by an experimentalist lying on the precise location
of the detector, i.e.,  $r_0 = r_d$. In principle, this 
would require that the scalar field were quantized with 
respect to a fiducial observer at $r_0$ (in which case $e^{-i\omega t}$ 
in Eq.~(\ref{normalmodespheric}) would be replaced by 
$e^{-i \tilde \omega \tau}$ with $\tau = \sqrt{f^{RN}(r_0)}\, t$). It is not 
difficult to see, however,
that ${\Gamma}^\alpha_{\omega_0 l {\rm m}}$ can be obtained directly from
Eq.~(\ref{P3}) with the following replacements:
$r_d \to r_0$ and $\omega \to \sqrt{f^{RN}(r_0)}\, \omega_0$.
[One can check this strategy in the Rindler wedge by obtaining
directly Eq.~(\ref{P2a'}) from Eq.~(\ref{P2'}).] 
In Fig.~(\ref{fig5accmv}) we plot ${\Gamma}^{\rightarrow}_{\omega_0 0 0}$
as a function of $\omega_0/m$ for observers at different locations.
(We recall that in four-dimensional spacetimes $c_0$ is dimensionless,
in contrast to the case where the spacetime is two-dimensional.) 

Now, let us define from Eq.~(\ref{P3}) the (scalar) normalized 
probability density 
\begin{equation}
{d{\cal P}^{\alpha}_{\omega l {\rm m}}}/{d V_d}  
\equiv 
{{\Gamma}^{\alpha}_{\omega l {\rm m}}({\bf x}_d)}/
{\int {\Gamma}^{\alpha}_{\omega l {\rm m}} ({\bf x}') d V' }
\; .
\label{spatialdensityBH}
\end{equation}
Here $ (d{\cal P}^{\alpha}_{\omega l {\rm m}}/dV_d) dV_d $  
is the probability that 
a  particle with quantum numbers ($\alpha, \omega, l, {\rm m})$ 
be found at some spatial point ${\bf x}_d = (r_d,\theta_d,\phi_d)$
in a proper volume $d V_d = d\lambda_d r_d^2 \sin \theta_d \, 
d\theta_d d\phi_d$,  where 
$d \lambda_d = \sqrt{h^{\rm RN} (r_d)} dr_d $. 
The proper radial distance can be integrated from the horizon 
$r=r_+$ as a function of the radial coordinate: 
\begin{eqnarray}
&& \lambda(r)
=  
\int_{r_+}^r dr' \sqrt{h^{\rm RN} (r')\, } 
\nonumber \\
&& = r \sqrt{f^{\rm RN}(r)} + 
            M \ln 
            \left[
            \frac{r + r \sqrt{f^{\rm RN}(r)} - M}{r_+ -M}
            \right] \; .
\end{eqnarray}
Notice that $\lambda = \lambda (r)$ is a regular function 
except in the case 
of extreme Reissner-Nordstrom black holes ($Q=M$).
From Eq.~(\ref{spatialdensityBH}), we obtain the radial-distance
probability density
\begin{equation}
\frac{d{\cal P}^{\alpha}_{\omega l {\rm m}}}{d \lambda_d}  
\equiv 
\frac{\sqrt{f^{\rm RN}(r_d)} |\psi^{\alpha}_{\omega l}(r_d)|^2}
{\int_0^{+\infty} d{\lambda_d}'  
\sqrt{f^{\rm RN}({r_d}')} |\psi^{\alpha}_{\omega l}({r_d}')|^2} \; .
\label{radialdensityBH}
\end{equation}
Note that here 
$(d{\cal P}^{\alpha}_{\omega l {\rm m}}/d \lambda_d) d \lambda_d $ 
is the probability of
detecting a particle with quantum numbers $(\alpha,\omega,l,{\rm m})$
inside a shell between $\lambda_d$ and $\lambda_d+ d\lambda_d$.  
In Fig.~(\ref{excitationdensityfig}), we use Eq.~(\ref{radialdensityBH}) 
to plot ${d{\cal P}^{\alpha}_{\omega l {\rm m}}}/{d \lambda_d}$
for particles with different  $\omega/m$ ratios (and $l=0$). We see that 
the smaller  the $\omega/m$ the closer
to the black hole horizon the particle will be in average.

\begin{figure}
\begin{center}
\mbox{\epsfig{file=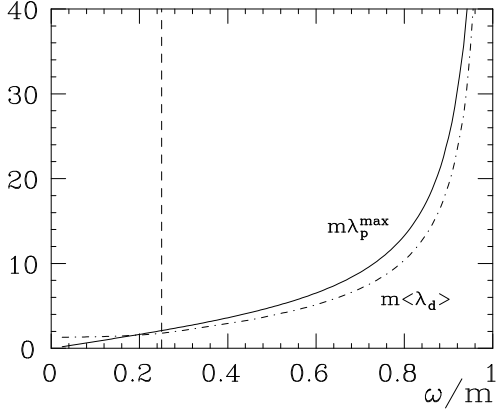,width=0.36\textwidth,angle=90}}
\end{center}
\caption{$\langle \lambda_d \rangle$ is plotted for $\alpha= \rightarrow$
and $l=0$ and compared with $\lambda^{\rm max}_p$. We see, indeed, that  
$\langle \lambda_d \rangle < \lambda^{\rm max}_p$ in the ``high-frequency''
regime $\omega > r_+^{-1}$
(i.e., at the right of the vertical broken line). (We have assumed here
$ mM = 2 $ and $Q=0$.)
}
\label{averagedetectionpositionfig}
\end{figure}

Next, in order to determine the mean  particle radial distance, 
we calculate numerically [see Fig.~(\ref{averagedetectionpositionfig})]
\begin{equation}
\langle \lambda_d \rangle  
\equiv 
\int_0^{+\infty} d{\lambda_d} \, \lambda_d \,
d {\cal P}^{\alpha}_{\omega l {\rm m}} / d\lambda_d \; .
\label{averageBH}
\end{equation}
Note that modes with $\omega/m \ge 1$ spreads over the whole 
space and thus $\langle \lambda_d \rangle$ diverges. This is qualitatively
different from the Rindler wedge case where the scattering 
potential grows unboundedly at the infinity what causes
the normal modes to vanish asymptotically. As a consequence, at the 
Rindler wedge, $\langle \rho_d \rangle$ is finite even for $\omega/m > 1$
[see Eq.~(\ref{average})].

Now, by repeating our General Relativistic analysis performed 
in the Rindler wedge case in a more compact way, we obtain 
that according to (our fiducial) static asymptotic observers a 
particle with  mass $m$  at some point 
$(r_p,\theta_p,\phi_p)$ with some four-velocity $u^\mu$ has 
total energy 
$\omega = m u^\mu (\partial_t)_\mu 
=
m \sqrt{f^{\rm RN}(r_p)} \sqrt{ 1 + |u^i u_i|\,} 
\geq m \sqrt{f^{\rm RN}(r_p)} $. (The equality holds when the particle 
is at rest.) By inverting this relation, we obtain that
a classical particle with mass $m$ and total energy $\omega$
(with $\omega < m$) will be at $r_p \leq r_p^{\rm max}$, where
\begin{equation}
r_p^{\rm max} = \frac{M[1+\sqrt{1-(1-\omega^2/m^2) (Q/M)^2}\,] }
                     {1-\omega^2/m^2} \; 
\label{rmaxRN}
\end{equation}
(and $r_p=r_p^{\rm max}$ only when the particle lies at rest).
Because the proper radial distance $\lambda= \lambda(r)$ is a 
growing monotonic function, 
$\lambda_p \leq \lambda_p^{\rm max} \equiv \lambda (r_p^{\rm max})$. 
In the ``high-frequency'' regime, $\omega \gg r_+^{-1}$,
we expect as before $\langle \lambda_d \rangle \leq \lambda^{\rm max}_p$
(see Sec.~\ref{Rw}).
This is indeed verified in Fig.~(\ref{averagedetectionpositionfig}).

\section{Detection of massive particles at Relativistic stars}
\label{RS}

Let us consider a static relativistic star with uniform
density $\alpha_0 = {\rm const.}$
The associated line element can be written  as in 
Eq.~(\ref{lineelementspheric}) (see Ref.~\cite{W}) with
\begin{eqnarray}
f^s(r) & = & \frac{1}{4}\left( 3\sqrt{1-\frac{2M}{R}} - 
           \sqrt{1-\frac{2Mr^2}{R^3}}
           \right)^2 \Theta (R-r)
\nonumber \\
& + & \left( 1-\frac{2M}{r} \right) \Theta (r-R)
\label{f^s}
\end{eqnarray}
and
\begin{eqnarray}
h^s(r)  & = &  (1-2m(r)/r)^{-1} 
\nonumber \\
        & = &  {\left(1-2M \frac{r^2}{R^3}\right)^{-1}}{\Theta (R-r)} 
\nonumber \\    
 & + & \left(1-\frac{2M}{r}\right)^{-1} {\Theta (r-R)} \;,
\label{h^s}
\end{eqnarray}
where
$
m(r) =  4\pi \int_0^r \alpha_0 {r'}^2 dr' \; .
$
We are using here the label ``$s$'' to denote quantities associated with
the {\em star} spacetime. In these coordinates, $r=R$ defines the star radius
and $M = (4\pi/3) \alpha_0 R^3$ is the star total mass 
(which, of course, differs from the star total {\em proper} mass 
because of the 
binding energy contribution). Note that for sake of stability, 
$R > R_c\equiv 9M/4$.  

\begin{figure}
\begin{center}
\mbox{\epsfig{file=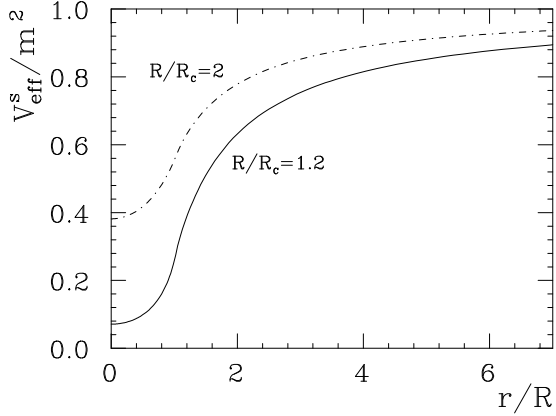,width=0.36\textwidth,angle=90}}
\end{center}
\caption{The scattering potential is plotted for different
$R/R_c$ values. We have assumed here $m M=2$. The denser the star,
the larger the influence of the gravitational field. The 
first eigenfrequencies for these potentials are listed
in Tab.~\ref{tabela}. }   
\label{Vstar}
\end{figure}

The positive-frequency solutions can be cast in
the same form given in Eq.~(\ref{normalmodespheric})
where $\psi^\alpha_{\omega l}$ is defined by 
Eq.~(\ref{radialequation}) with $f(r)$ and $h(r)$ given in
Eqs.~(\ref{f^s}) and (\ref{h^s}), respectively. 
The scattering potential can be obtained by using 
Eqs.~(\ref{f^s}) and (\ref{h^s}) in Eq.~(\ref{Veff}):
\begin{eqnarray}
& &
 V^s_{\rm eff}(r) 
= 
\left[ 
\frac{M}{R^3} 
  \left( 
   \frac{-9 F(R)}{2} + \frac{9}{2}\sqrt{\frac{F(R)}{h^s(r)}} 
   - \frac{1}{h^s(r)}
  \right)
\right.
\nonumber \\              
& &
\left.
+\frac{1}{4} \left( \frac{l(l+1)}{r^2} + m^2 \right)
\left( 9 F(R) + \frac{1}{h^s(r)} -6 \sqrt{\frac{F(R)}{h^s(r)}} \right)
\right]
\nonumber \\ 
& & 
\times \Theta(R-r)
\! +\!
f^s(r)\! \left( \frac{2M}{r^3}\! +\! \frac{l(l+1)}{r^2}\! +\! m^2 \right)\! 
\Theta(r-R)
\label{Vs}
\end{eqnarray}
where $F(R) \equiv 1-2M/R$.
At the star surface the potential has a discontinuity:
\begin{eqnarray}
\lim_{r \to R_-} V^s_{\rm eff}(r) 
& = &
F(R)  \left(\frac{l(l+1)}{R^2}-\frac{M}{R^3} + m^2 \right)  \; ,
\nonumber \\
\lim_{r \to R_+} V^s_{\rm eff}(r)
& = &
F(R)  \left(\frac{l(l+1)}{R^2}+\frac{2M}{R^3} + m^2 \right) \; . 
\nonumber
\end{eqnarray}
In Fig.~\ref{Vstar} we plot the scattering potential 
for different star parameters. (Note that the discontinuity of 
$V^S_{\rm eff}$ at $r/R=1$ is very small for the parameters chosen 
in Fig.~\ref{Vstar} to be self-evident.)
It is interesting to observe that the gravitational field of the Earth
is too faint to significantly influence the scattering potential for 
an electron-mass particle ($m=0.5\, {\rm MeV}$). At the center
where the influence would be maximum, we would have for 
$M= 6 \cdot 10^{27}{\rm g}$ (and $l=0$) that 
$V^s_{\rm eff}(r) = m^2 (1-10^{-48})$ which should be compared with  
$\left. V^s_{\rm eff}(r)\right|_{M=0} = m^2 $ obtained in the absence of 
gravity. Notwithstanding, for quasi-extreme stars ($R \approx R_c$) 
the scattering potential can be quite affected by the gravitational 
field~(see Fig.~\ref{Vstar}). Indeed it is possible to devise situations 
where the gravitational field of a star would be of some importance for 
particle-physics processes~\cite{VM3}.
\begin{figure}
\begin{center}
\mbox{\epsfig{file=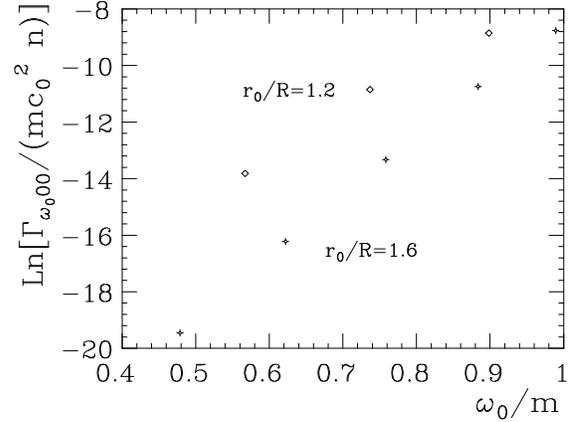,width=0.36 \textwidth,angle=90}}
\end{center}
\caption{We plot the detection rate ${\Gamma}_{\omega_0 00}$ 
for the eigenfrequencies $\omega_0/m < 1$ assuming observers at 
different points $r_0$. We note that the damping in the detection 
rate is steeper for observers at weak gravitational potentials.
(Here $mM=2$ and $R/R_c=1.2$.)
}   
\label{fig8bccmv}
\end{figure}

The normal modes with $\omega < m$ will be 
bounded states and $\omega_{\rm min} \equiv \omega_1$ will be the lowest 
eigenvalue of the discrete frequency spectrum.
Moreover, there will exist only
one set of normalizable solutions and so we will suppress
the ``$ \alpha$'' label hereafter. 
By defining 
$d{\tilde r} = (2M)^{-1} \sqrt{h^s(r)/f^s(r)} dr $, we can write
Eq.~(\ref{radialequation}) as  
\begin{equation}
-\frac{d^2 \psi_{\omega l}}{d {\tilde r}^2} + 
4 M^2 V^s_{\rm eff} [{\tilde r}(r)] \psi_{\omega l} 
= 4 M^2 \omega^2 \psi_{\omega l} \; .
\label{radialequation3}
\end{equation}
The function $\tilde r (r)$ is a growing monotonic function.
At the star core (and choosing properly the integration constant)
we have
${\tilde r}|_{r\approx 0} \approx (r/M)/(3 \sqrt{F(R)\,} -1 )$
and the scattering potential becomes
\begin{equation}
\left. V^s_{\rm eff} (\tilde r) \right|_{\tilde r \approx 0} 
 \approx \frac{l(l+1)}{4 M^2 {\tilde{r}}^2} + C^2 + {\cal O} (\tilde{r}) \; ,
\label{Veffsatthecenter}
\end{equation}
where 
\begin{equation}
C^2  =  \frac{G(R)^2}{2 M^2} 
\left[ 
\frac{(mM)^2}{2} 
+ \frac{M^3  l(l+1)}{R^3 G(R)} 
+ \frac{M^3}{R^3 G(R)}
- \frac{M^3}{R^3}
\right]
\end{equation}
is a positive-definite constant and
$G(R) \equiv -1+3 \sqrt{F(R)} $.
Thus, close to the star center, $\psi_{\omega l} $ is proportional to 
$2 M \nu \tilde{r} j_l (2 M \nu \tilde{r}) $
(since the solutions proportional to 
$2 M \nu \tilde{r} n_l (2 M \nu \tilde{r})$
are non-normalizable), where $j_l(x)$ is
the spherical Bessel function and 
 $\nu \equiv \sqrt{\omega^2 - C^2}$. 
Moreover, far away from the star,  $V^s_{\rm eff}(r) \approx m^2$.
Hence
\begin{equation}
     \psi_{\omega l} (\tilde{r}) \approx 
        A_{\omega l}
         \left\{ 
          \begin{array}{lc}
           B_{\omega l} 
           2 M \nu \tilde{r} j_l (2 M \nu \tilde{r}) & (\tilde r \approx 0) \\ 
           C_{\omega l} e^{-2iM \tilde \omega \tilde r} + 
           e^{2iM \tilde \omega \tilde r} & (\tilde r \gg 1) 
          \end{array} 
        \right.   
   \label{51b}
\end{equation} 
where we recall that $\tilde \omega \equiv \sqrt{\omega^2 - m^2}$. 
For $\omega \ge m$, the modes are asymptotically free states
and they can be normalized through Eqs.~(\ref{KG1})-(\ref{KG2}). Indeed
we find $A_{\omega l} = (2\sqrt{\omega \tilde \omega})^{-1}$ and 
$|C_{\omega l}| = 1 $. For $\omega_i < m$, the normal modes fade exponentially
and $|C_{\omega_i l}| = 0 $, where we are using here latin indeces 
$i,j,\ldots $ to lable  the discrete eigenfrequencies.
The bounded states should be 
Klein-Gordon orthonormalized by imposing Eqs.~(\ref{KG1})-(\ref{KG2})  
except for the fact that in this case $\delta_{A A'} \equiv \delta_{l l'} 
\delta_{{\rm m} {\rm m}'} \delta_{\omega_i \omega_j}$. Thus, we find
\begin{equation}
\int_0^{+\infty} d\tilde r \; |\psi_{\omega_i l} (\tilde{r})|^2
                            = 
\frac{\pi}{4 M \omega_i^2} \; .
\end{equation} 
The valid set of discrete eigenfrequencies $\omega_i$ 
is determined numerically~\cite{G}. Shortly, the strategy consists in
evolving solutions with $\psi_{\omega_i l} (0) =0$ [see Eq.~(\ref{51b})]
(and arbitrary  $\psi_{\omega_i l}' (\tilde r)|_{\tilde r = 0} = {\rm const.}$)
 and search for the $\omega_i$'s such that far enough from the star
$\psi_{\omega_i l} \sim e^{-2M\sqrt{m^2 - \omega_i^2} \, \tilde r}$ as ruled by
Eq.~(\ref{51b}) for $\omega_i < m$. In Tab.~\ref{tabela} we present the
lowest  eigenfrequencies 
$\{\omega_1, \omega_2, \omega_3, \omega_4 \}$ (normalized by $m$)
for some star parameters. For sake of comparison, we also show the
minimum energy value $\omega^{\rm class}_{\rm min}$ for a {\em classical} 
particle obtained according to General Relativity when it 
is static at the star center. Namely, 
$\omega^{\rm class}_{\rm min} = 
 m \sqrt{f(r)}|_{r=0} =  (m/2) (3\sqrt{1-2M/R}-1)$. 
In the ``high-frequency'' regime, $\omega^{\rm class} \gg M/R^2 $, 
where the behavior of classical  and quantum particles can be compared, 
we expect  $\omega_i \geq \omega^{\rm class}_{\rm min}$
because of the extra intrinsic ``kinetic energy'' of quantum 
origin (see Tab.~\ref{tabela}).
\begin{table}
\begin{tabular}{|c||c|c|c|c|c|} 
     & $\omega^{\rm class}_{\rm min}/m$ 
     & $\omega_1/m$ 
     & $\omega_2/m$ 
     & $\omega_3/m$ 
     & $\omega_4/m$ \\ \hline
$mM =2, R/R_c=1.2$ 
	&   0.264
	&   0.351      
	&   0.456    
	&   0.556      
	&   0.648  \\ \hline
$mM =2, R/R_c=2.0$ 
	&   0.618
	&   0.679
        &   0.757
        &   0.826 
        &   0.874
\end{tabular}
\vskip 0.5 truecm
\caption{We list the lowest eigenfrequencies $\{ \omega_1,..., \omega_4 \}$ 
for different star parameters. The lowest eigenfrequency 
$\omega_1$ is to be compared with the minimum
classical energy $\omega^{\rm class}_{\rm min}$. We see that 
$\omega^{\rm class}_{\rm min} \leq \omega_1$, as expected, in the 
``high-frequency'' regime $\omega^{\rm class} \gg M/R^2$. By assuming that
$mM=2$, 
then $M/R^2 =0.07\, m$ and $M/R^2 =0.02\, m$  for $R/R_c=1.2$ and $R/R_c=2$, 
respectively.}
\label{tabela}
\end{table}

The massive scalar field can be cast as in Eq.~(\ref{Phisphericalsymmetric})
\begin{equation} 
\hat \Phi(x) = 
\sum_{l=0}^{+\infty} 
\sum_{{\rm m}=-l}^{l} 
\sum_{\omega} \!\!\!\!\!\!\!\!\! {\int} 
\;\;\;[ \hat a_{\omega l {\rm m}} u_{\omega l {\rm m}} (x) + {\rm H.c.}] \;,
\label{Phisphericalsymmetric2}
\end{equation}
where  the summation in $\alpha$ was suppressed and
we integrate over the free states ($\omega \geq m$) and 
sum over the eigenfrequencies of the bounded states ($\omega_i < m$). 
Here 
\begin{equation}
     \left[ 
     \hat a_{\omega l {\rm m} }, {\hat a^{\dagger}_{\omega' l' {\rm m}'}} 
     \right] = \delta_{A A'} \;,
   \label{commutationrelation3}
\end{equation}
where 
$
\delta_{A A'} \equiv 
\delta_{l l'} \delta_{{\rm m} {\rm m}'} \delta(\omega - \omega')
$
and  
$
\delta_{A A'} \equiv 
\delta_{l l'} \delta_{{\rm m} {\rm m}'}  \delta_{\omega \omega'}
$
for free and bounded states, respectively.

Next we couple [see Eq.~(\ref{S})] the massive scalar field to an
Unruh-DeWitt detector characterized by the density of states~(\ref{beta}).
The amplitude (at the tree level) associated with the detection of a bounded 
state with quantum numbers $ \omega_i, l$ and ${\rm m}$ is 
$
{\cal A}_{\omega_i l {\rm m}}^{\rm det} 
= 
\langle 0
| \otimes \langle E | \hat S_I | E_G  \rangle \otimes | 
\omega_i l m\rangle
$. 
The detection rate (detection probability per detector proper time) 
is, thus,
\begin{eqnarray}
{\Gamma}_{\omega_i l {\rm m}} 
&=& 
\frac{1}{s_d^{\rm tot}} \int_0^{+\infty} dE \;\beta_{E_0} (E) 
|{\cal A}_{\omega_i l {\rm m}}^{\rm det}|^2 
\nonumber \\
&=&    
2 c_0^2 n \sqrt{f^s( r_d )} 
\frac{|\psi_{\omega_i l} ( r_d)|^2}{r_d^2} 
|Y_{l {\rm m}} (\theta_d, \varphi_d)|^2 \; ,
\label{P3s}
\end{eqnarray}
where again
the detector selectivity was chosen such that
$\langle E | m(0) | E_G  \rangle \equiv 1$ and
we have tuned the energy gap  of the detector 
as usually:
$E_0 = \omega_i/\sqrt{f^s(r_d)} $. 
[Notice the resemblance of Eqs.~(\ref{P3s}) and~(\ref{P3}), 
where we recall that  ${\rm dim} (\psi_{\omega_i l}/\psi^\alpha_{\omega l}) 
= {\rm dim} (\sqrt{\omega})$.]
(The detection of free states follows as previously [see, e.g., 
Eq.~(\ref{P2})].)
\begin{figure}
\begin{center}
\mbox{\epsfig{file=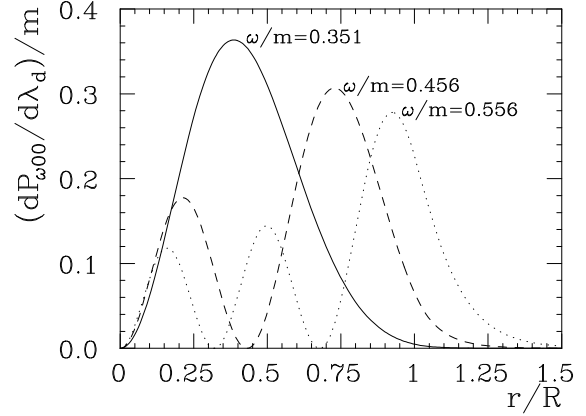,width=0.36\textwidth,angle=90}}
\end{center}
\caption{The probability density 
$(d{\cal P}_{\omega 0 0}/d\lambda_d)/m$ is plotted for 
the three lowest eigenfrequencies when $mM=2$ and $R/R_c=1.2$
The smaller the $\omega/m$ ratio, the closer to the star
center the particle is.}
\label{excitationdensityfigs}
\end{figure}
As in the Reissner-Nordstrom case, we can calculate the detection 
rate ${\Gamma}_{\omega_0 l m}$ in the special 
case where the massive state $|\omega_0 l {\rm m}\rangle$ is
defined  by an experimentalist lying on the detector, 
as shown in Fig.~\ref{fig8bccmv}. 

\begin{figure}
\begin{center}
\mbox{\epsfig{file=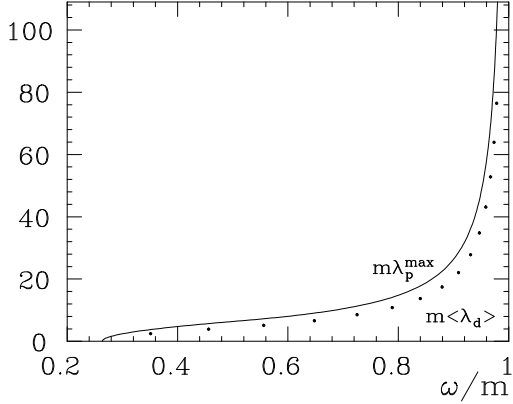,width=0.36\textwidth,angle=90}}
\end{center}
\caption{$\langle \lambda_d \rangle$ is plotted for various 
eigenfrequencies assuming $l=0$,  $mM =2$ and $R/R_c=1.2$.
We note that $\langle \lambda_d \rangle < \lambda^{\rm max}_p$, 
as expected, in the ``high-frequency'' regime
$\omega^{\rm class}/m \gg (M/R^2)/m$. 
(Here $\omega^{\rm class}_{\rm min}/m = 0.264$,
$\omega_1/m = 0.351$ and
$(M/R^2)/m = 0.07$.) }
\label{averagedetectionpositionfigs}
\end{figure}

By using Eq.~(\ref{P3s}), we can define a normalized probability density 
${d{\cal P}_{\omega_i l {\rm m}}}/{d V_d}$ analogously to 
Eq.~(\ref{spatialdensityBH}), where 
$dV_d = d\lambda_d r_d^2 \sin \theta_d d \phi_d$
and $d\lambda_d =  \sqrt{h^s(r_d)} dr_d$.
The proper radial distance can be integrated from the center of the star
as a function of the radial coordinate leading to
\begin{eqnarray}
& & \lambda  =  \frac{\arcsin[\sqrt{2M r^2/R^3}]}{\sqrt{2M/R^3}} 
                  \Theta[R-r]  + \!
                  \left\{ 
                  \frac{\arcsin [\sqrt{2M/R}]}{\sqrt{2M/R^3}}
                  \right. 
\nonumber \\                
& &               + M\ln \left[
                        \frac{r + r \sqrt{1-2M/r} - M}
                                   {R + R \sqrt{1-2M/R} - M}
                       \right] +r\sqrt{1-\frac{2M}{r}\,} 
\nonumber \\
& &             \left.  
                - R \sqrt{1- \frac{2M}{R}\,}
            \right\}  \Theta[r-R] \; .          
\label{properdistances}
\end{eqnarray}
The radial-distance probability density 
$d{\cal P}_{\omega_i l {\rm m}}/d \lambda_d$ is then calculated
analogously to Eq.~(\ref{radialdensityBH}). In 
Fig.~(\ref{excitationdensityfigs}), we plot 
$d{\cal P}_{\omega_i l {\rm m}}/d \lambda_d$
for particles with different  $\omega_i/m$ ratios.
The mean particle radial distance $\langle \lambda_d \rangle$ 
is defined analogously to Eq.~(\ref{averageBH}) and was  
calculated numerically as shown in 
Fig.~(\ref{averagedetectionpositionfigs}).

Now, by taking a procedure analogous to the one used in the black hole case 
[see Eq.~(\ref{rmaxRN})], we find that a classical 
particle with total energy $\omega < m$ (as measured by static asymptotic 
observers) will be at $r_p\leq r_p^{\rm max}$, where
\begin{eqnarray}
r_p^{\rm max} 
& = & 
\sqrt{
\frac{R^3}{2M}\! 
\left[ 
1-\left( 3 \sqrt{F(R)} -2\frac{\omega}{m} \right)^2
\right]
}
\Theta \left[ \sqrt{F(R)}\! -\! \frac{\omega}{m} \right]
\nonumber \\
& + &
\frac{2M}{1-(\omega/m)^2} \Theta \left[\frac{\omega}{m}-\sqrt{F(R)} \right]\; .
\end{eqnarray}
Because the proper radial distance $\lambda= \lambda (r)$ is 
again a growing monotonic
function [see Eq.~(\ref{properdistances})], we have once more
$
\lambda_p \leq \lambda_p^{\rm max} \equiv \lambda (r_p^{\rm max})  \; .
$
This is compared with
$\langle \lambda_d \rangle $ in Fig.~(\ref{averagedetectionpositionfig}).
We see that $\langle \lambda_d \rangle < \lambda_p^{\rm max}$,
as expected, in the ``high-frequency'' regime, 
$\omega^{\rm class} \gg M/R^2$.

\section{Conclusions}
\label{conclusions}

The detection rate for massive particles with total energy $E<mc^2$ 
was calculated for the Rindler wedge, black holes and star spacetimes.
The mean particle positions were calculated and shown to be in
qualitative agreement with General Relativity predictions in the
``high-frequency'' regime. 
The observation process is naturally taken into account in the formalism
by considering Unruh-DeWitt-like detectors. In this way, we clearly define
what we mean by ``observing a particle''. 

Although it is possible, in principle, to measure particles
satisfying Eq.~(\ref{fundamental}) even at Earth, 
these events  are so rare in such weak gravitational
fields that do not have to be  considered for
practical purposes. This can be seen from the 
analysis of the scattering potential [see discussion 
below Eq.~(\ref{Vs})].

Notwithstanding the consideration of massive particles
with {\em arbitrary small} total energy  
is just fundamental for a  comprehensive understanding of some
phenomena occurring in highly curved spacetimes (e.g., black holes)
and accelerated frames. 
For example, the weak decay of uniformly accelerated protons in the Minkowski 
vacuum which is described by the  $p \to n \; e^+ \; \nu$ channel
(at the tree level) in an inertial frame was shown to be representable 
by the combination of the following three channels:
$p \; e^-      \to n \;  \nu \;,$ 
$p \; \bar \nu \to n \;  e^+ \;$ 
and 
$p \; e^-\; \bar \nu \to n \;,$
in the proton coaccelerated frame. According to this description, 
the proton would decay by the absorption of a Rindler $e^-$  and/or 
$\bar \nu$ 
from the Fulling-Davies-Unruh thermal bath ``attached'' to the proton 
proper frame. However, in order that inertial and 
noninertial frame descriptions lead to the same observables (e.g. 
 proper lifetime), we emphasize that the energy
spectrum of the (massive) Rindler $e^-$'s and $e^+$'s must be 
$\omega \in [0,+\infty)$ (despite Rindler $e^-$'s and $e^+$'s 
have mass $m \approx 0.5\; {\rm MeV}$, as usually)~\cite{MV1}. 
Perhaps an even more extreme case concerns the importance
of {\em zero-energy} particles to understand the
radiation emitted from uniformly accelerated charges according 
to comoving observers. Namely, the emission of a (usual) Minkowski
photon with transverse momentum ${\bf k}_\bot$ 
from a uniformly accelerated charge as described by inertial 
observers corresponds, according to coaccelerated observers, to either
the {\em absorption from} or the {\em emission to} the Fulling-Davies-Unruh
thermal bath of a zero-energy photon with the same transverse momentum 
${\bf k}_\bot$.  For obvious reasons, in Ref.~\cite{HMS} the photons 
considered were supposed to be massless but the same conclusion in terms
of zero-energy particles would hold even if we considered that 
the photons were {\em massive}. 
 
\begin{flushleft}
{\bf{\large Acknowledgements}}
\end{flushleft}
We are indebted to Professor L. Parker for reading the manuscript.
We are grateful to Dr. A. Gammal for calling our attention to Ref.~\cite{G}.
One of us (D.V.) is thankful to Drs. A. Higuchi and B. Kay for conversations
on this issue. We acknowledge Funda\c c\~ao de Amparo \`a Pesquisa do Estado 
de S\~ao Paulo for supporting J.C. (fully) and D.V. (partially)
and Conselho Nacional  de Desenvolvimento  Cient\'\i fico e  
Tecnol\'ogico for supporting G.M. (partially). This work was  supported
in part by the US National Science Foundation under Grant: No PHY-0071044.

\end{document}